# Mathematical Model of Shock Waves

Alexei Krouglov

*1550  16th Ave., Bldg. F, 2nd Floor, Richmond Hill, Ontario L4B 3K9, Canada*

Email: Alexei.Krouglov@siconvideo.com

# ABSTRACT

Presented here is the mathematical model describing the phenomenon of shock waves. The underlying concept is based on the time-space model of wave propagation.

## 1. Concept behind the Phenomenon

The first assertion is that a wave is caused by the difference between energy's value at some point and "average" energy's values (coined as an energy's level) in the "neighborhood" of this point. That difference is called energy's discrepancy at this point, which is wavily propagated both in space and in time. Besides that a wave travels with a finite speed (see [1]).

The second assertion is that additionally to wavy propagation the energy's discrepancy spreads directly as well. But direct spreading decelerates quickly whereas wave may propagate on a big distance.

The conclusion is that when speed of direct spreading of energy's discrepancy exceeds the speed of wave propagation, the wave acquires an additional energy while travels in space. And we have to apply the wave's attenuation as well.

Therefore an accumulation of energy's discrepancy by traveling waves creates the phenomenon, which received a name of shock waves.

## 2. Model Description

Here I describe the one-dimensional case.

I assume that at time $t = t_0$ in point $x = x_0$ we have a positive energy's discrepancy,

$$\Delta \Phi_0(t_0, x_0) > 0$$

According to [1], this discrepancy propagates wavily in space with a finite speed $c > 0$. While traveling the wave attenuates. Therefore, at time $t = t_1 > t_0$ when a discrepancy reaches the point $x = x_1$, where

$$x = x_0 + (t_1 - t_0) \cdot c > x_0$$

it will have the amplitude

$$\Delta\Phi_0(t_1, x_1) = \Delta\Phi_0(t_0, x_0) - \int_{x_0}^{x_1} r(x)dx < \Delta\Phi_0(t_0, x_0) \tag{1}$$

where $r(x) > 0$ represents the losses of energy's discrepancy at points $x \in [x_0, x_1]$.

On the other hand, energy's discrepancy spreads directly in interval $[x_0, x_2]$ quickly attenuating. I assume that the energy's discrepancy spreads with a speed $v > c > 0$.

Thus the discrepancy accumulates, and the accumulated value is

$$e(x) = \begin{cases} \int_{x_0}^{x} m(\bar{x})d\bar{x} & \text{for } x \leq x_2 \\ \int_{x_0}^{x_2} m(x)dx & \text{for } x > x_2 \end{cases} \tag{2}$$

where $m(x) > 0$ is a positive monotonically decreasing function.

Hence the overall energy's discrepancy at time $t = t_1$ in point $x = x_1$ (where it holds $x_0 < x_1 < x_2$) is

$$\Delta\Phi(t_1, x_1) = \Delta\Phi_0(t_0, x_0) + \int_{x_0}^{x_1} (m(x) - r(x))dx \tag{3}$$

Therefore when $m(x) > r(x)$ for $x \in [x_0, x_1]$ it holds the inequality,

$$\Delta\Phi(t_1, x_1) > \Delta\Phi_0(t_0, x_0)$$

and the energy's discrepancy grows for the traveling wave.

The foregoing description explains the phenomenon of shock waves.